\documentclass[twocolumn,bibnotes,showpacs,amsmath,amssymb,aps,prl]{revtex4-2}

\usepackage{graphicx}
\usepackage{amsmath}
\usepackage{amssymb}
\usepackage{color}
\usepackage{comment}
\usepackage[normalem]{ulem}
\usepackage{bm}

\usepackage[colorlinks=true,linkcolor=blue,citecolor=blue]{hyperref}

\begin{document}
\title{Impact of tiny Fermi pockets with extremely high mobility on the Hall anomaly\\
in the kagome metal CsV$_3$Sb$_5$}
\author{S.~Liu$^1$}\email{s.liu@qpm.k.u-tokyo.ac.jp}
\author{M.~Roppongi$^1$}
\author{M.~Kimata$^{2,3}$}
\author{K.~Ishihara$^1$}
\author{R.~Grasset$^4$}
\author{M.~Konczykowski$^4$}
\author{B.~R.~Ortiz$^5$}
\author{S.~D.~Wilson$^6$}
\author{K.~Yoshimi$^7$}
\author{T.~Shibauchi$^1$}\email{shibauchi@k.u-tokyo.ac.jp}
\author{K.~Hashimoto$^1$}\email{k.hashimoto@edu.k.u-tokyo.ac.jp}
\affiliation{
$^1$Department of Advanced Materials Science, University of Tokyo, Kashiwa, Chiba 277-8561, Japan\\
$^2$Institute for Material Research, Tohoku University, Sendai, Miyagi 980-8577 464-8602, Japan\\
$^3$Advanced Science Research Center, Japan Atomic Energy Agency, Tokai, Ibaraki 319-1195, Japan\\
$^4$Laboratoire des Solides Irradi{\' e}s, CEA/DRF/IRAMIS, Ecole Polytechnique, CNRS, Institut Polytechnique de Paris, F-91128 Palaiseau, France\\
$^5$Materials Science and Technology Division, Oak Ridge National Laboratory, Oak Ridge, TN 37831, USA\\
$^6$Materials Department, University of California Santa Barbara, Santa Barbara, CA 93106, USA\\
$^7$Institute of Solid Physics, University of Tokyo, Kashiwa, Chiba 277-8581, Japan
}

\begin{abstract}
The kagome metal CsV$_3$Sb$_5$ exhibits an unusual charge-density-wave (CDW) order, where the emergence of loop current order that breaks time-reversal symmetry (TRS) has been proposed. A key feature of this CDW phase is a non-monotonic Hall effect at low fields, often attributed to TRS breaking. However, its origin remains unclear.
Here, we conduct comprehensive magnetotransport measurements on CsV$_3$Sb$_5$ and, through mobility spectrum analysis, identify the formation of tiny Fermi pockets with extremely high mobility below the CDW transition. 
Furthermore, electron irradiation experiments reveal that the non-monotonic Hall effect is significantly suppressed in samples with reduced mobility, despite no substantial change in the electronic structure. 
These results indicate that the non-monotonic Hall effect originates from these tiny Fermi pockets with high mobility carriers rather than anomalous Hall mechanisms, providing new insights into understanding the Hall anomaly in this kagome system.
\end{abstract}

\maketitle

The recently discovered kagome metals $A$V$_3$Sb$_5$ ($A$\,=\,K,\,Rb,\,Cs) have garnered considerable attention as a platform for exploring various nontrivial quantum phases predicted in kagome lattice systems\,\cite{ortiz2020}. This family features a perfect kagome lattice formed by V atoms, resulting in a unique band structure with Dirac cones and van Hove singularities near the Fermi energy, which are characteristic of ideal kagome lattices\,\cite{tan2021,Kang2022,Hu2022}. Theoretically, these van Hove singularities are expected to induce instabilities that drive unconventional charge-density-wave (CDW) and superconducting states\,\cite{Lin2021,Wu2021,ortiz2020}. Indeed, these materials undergo a CDW transition at $T_{\rm CDW}=78$--103\,K\,\cite{ortiz2019,Ortiz2021PRM,ortiz2020}, characterized by lattice modulations with star-of-David and/or inverse star-of-David (trihexagonal) patterns\,\cite{Kautzsch2023,Frassineti2023,Stahl2022,Xiao2023}, followed by a superconducting transition at $T_{\rm c}=0.9$--2.5\,K\,\cite{ortiz2020,Ortiz2021PRM,yin2021}.

In these kagome materials, several exotic phenomena, such as time-reversal symmetry (TRS) and rotational symmetry breakings, have been reported in relation to the CDW order. For instance, various experiments, including muon spin relaxation ($\mu$SR)\,\cite{Mielke2022,yu2021}, Kerr effect\,\cite{saykin2023,hu2023}, scanning tunneling microscopy (STM)\,\cite{Jiang2021,Zhao2021,Wang2021}, and magnetic torque measurements\,\cite{Asaba2024}, have pointed out the emergence of TRS breaking associated with the CDW order, and its possible connection to loop current order has been discussed\,\cite{Tazai2023}. Intriguingly, in this system, a non-monotonic Hall effect has been observed at low fields below $T_{\rm CDW}$\,\cite{yang2020,yu2021AHE,wang2023}. 
The origin of this non-monotonic Hall effect has been discussed in terms of various anomalous Hall mechanisms, including skew scattering caused by spin clusters\,\cite{yang2020}, Berry curvature arising from a topological band structure\,\cite{Fu2021,Chapai2023}, and spontaneous magnetization due to loop current order\,\cite{Jiang2021,Guo2022,Asaba2024}. 
In addition to these mechanisms, the band structure associated with van Hove singularities has also been considered as a possible origin of the Hall anomaly\,\cite{Koshelev2024}. 
However, despite extensive experimental and theoretical investigations, the origin of the Hall anomaly remains an open question.

Quantum oscillation experiments in this kagome system have shown that the Fermi surfaces are reconstructed at the CDW transition, resulting in the formation of multiple small Fermi surfaces with light effective masses\,\cite{ortiz2021,Broyles2022,Tan2023,yu2021AHE,Fu2021,Gan2021,Chen2022,Zhang2022,Shrestha2022,Chapai2023}. 
Importantly, multiband systems with small Fermi surfaces hosting high-mobility carriers often exhibit non-monotonic Hall effects at relatively low magnetic fields.
However, in this kagome system, a comprehensive analysis of magnetotransport properties that fully incorporates the multiband effect is limited, largely due to the complexity of accounting for multiple Fermi surfaces.
In this study, we conduct extensive magnetotransport measurements on CsV$_3$Sb$_5$ to clarify the origin of the non-monotonic Hall effect, with careful consideration of the multiband effect.
Using mobility spectrum ($\mu$-spectrum) analysis, which does not requires assumptions regarding the number or types of carriers\,\cite{mcclure1958,Beck1987,Antoszewski2012,Huynh2014}, we demonstrate that the non-monotonic Hall effect originates from small Fermi pockets with high-mobility carriers.
Furthermore, we confirm from the investigation of impurity effects that the non-monotonic Hall effect does not originate from anomalous Hall mechanisms.

\begin{figure*}[htbp]
    \includegraphics[width=0.8\linewidth]{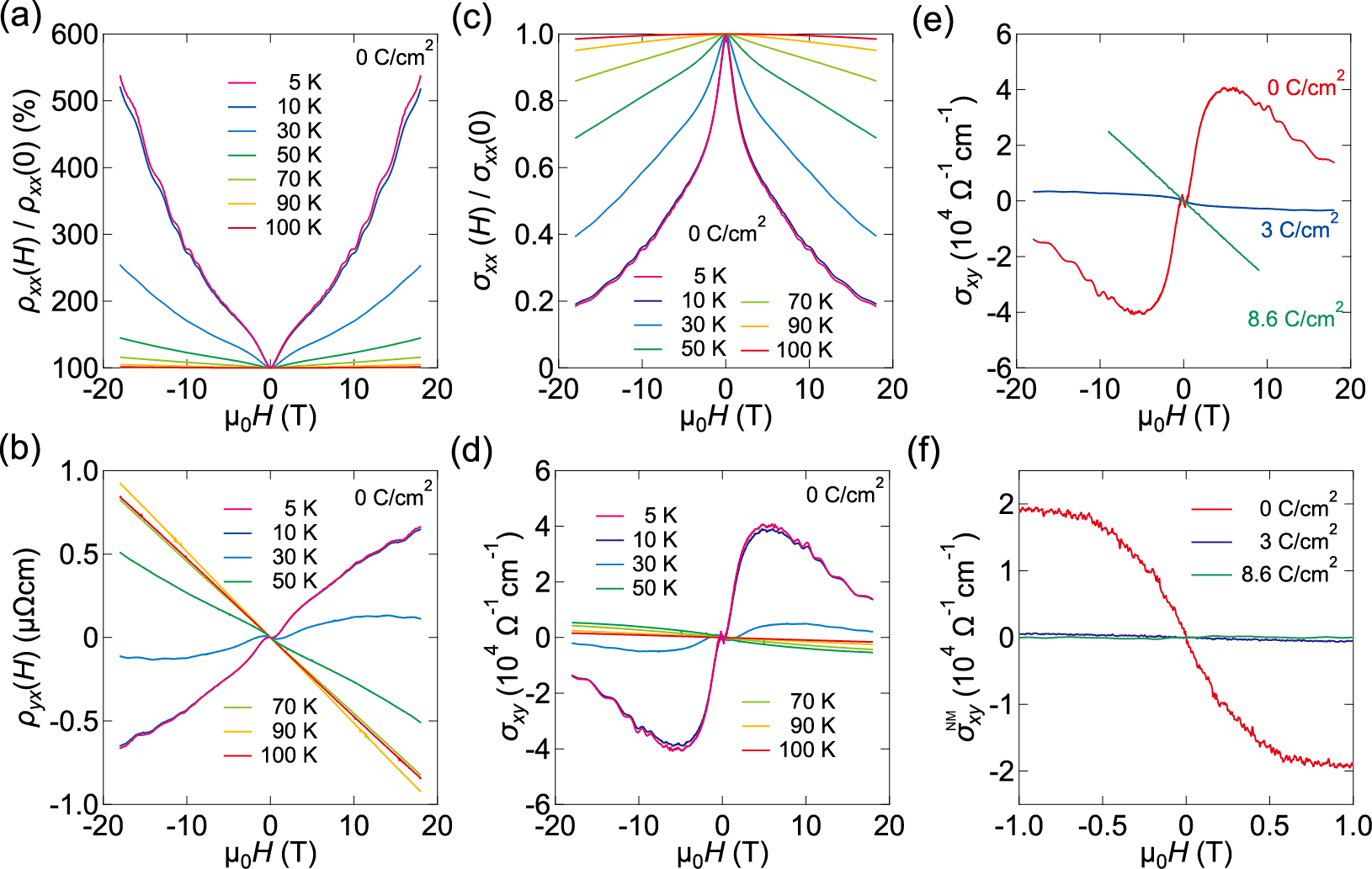}
    \caption{(a) Normalized MR of pristine CsV$_3$Sb$_5$ as a function of magnetic field applied along the $c$-axis, measured at various temperatures. (b) Hall resistivity of pristine CsV$_3$Sb$_5$ as a function of field along the $c$-axis, measured at various temperatures. (c) Normalized MC of pristine CsV$_3$Sb$_5$, measured at various temperatures. (d) Hall conductivity of pristine CsV$_3$Sb$_5$ as a function of field, measured at various temperatures. (e) Hall conductivity of pristine and electron-irradiated CsV$_3$Sb$_5$ samples as a function of field, measured at 5\,K. (f) Hall conductivity with the ordinary Hall component (proportional to the field) subtracted at 5\,K for pristine and irradiated CsV$_3$Sb$_5$ samples as a function of field.}
    \label{Fig1}
\end{figure*}

High-quality single crystals of CsV$_3$Sb$_5$ were synthesized using the self-flux growth method\,\cite{ortiz2020}. Magnetotransport measurements were conducted via a conventional five-probe method, using an 18-T superconducting magnet at the Institute for Materials Research, Tohoku University, and a Physical Property Measurement System (PPMS) up to 9\,T from Quantum Design.
Electron irradiation at an incident energy of 2.5\,MeV was carried out at the SIRIUS Pelletron accelerator, operated by the Laboratoire des Solides irradi\'{e}s (LSI) at \'{E}cole Polytechnique. 
The $\mu$-spectrum analysis was performed using the Gpufit package in Python\,\cite{gpufit}, and final fitting confirmation was conducted within the 2DMAT framework\,\cite{2DMAT}.

Figure\,\ref{Fig1}(a) shows the field dependence of the magnetoresistivity (MR) normalized by its zero-field value, $\rho_{xx}(H)/\rho_{xx}(0)$, measured up to $\pm$18\,T at various temperatures for the pristine CsV$_3$Sb$_5$ sample. As the temperature decreases, the MR increases rapidly, and Shubnikov-de Hass (SdH) quantum oscillations become pronounced. Figure\,\ref{Fig1}(b) presents the Hall resistivity $\rho_{yx}(H)$ as a function of field up to $\pm$18\,T for the pristine sample. At 100\,K, $\rho_{yx}$ is negative, but as the temperature decreases, its slope gradually decreases and eventually turns positive below $\sim$30\,K. Additionally, a non-monotonic Hall effect is observed in the low-field region below $\sim$1\,T. These results are consistent with previous studies\,\cite{yu2021AHE,ortiz2021}. Figures\,\ref{Fig1}(c) and \ref{Fig1}(d) show the field dependence of the normalized magnetoconductivity (MC), $\sigma_{xx}(H)/\sigma_{xx}(0)$, and the Hall conductivity, $\sigma_{xy}(H)$, for the pristine sample, respectively. The Hall anomaly in the low-field region is clearly visible at low temperatures.

\begin{figure*}[htbp]
    \centering
    \includegraphics[width=0.8\linewidth]{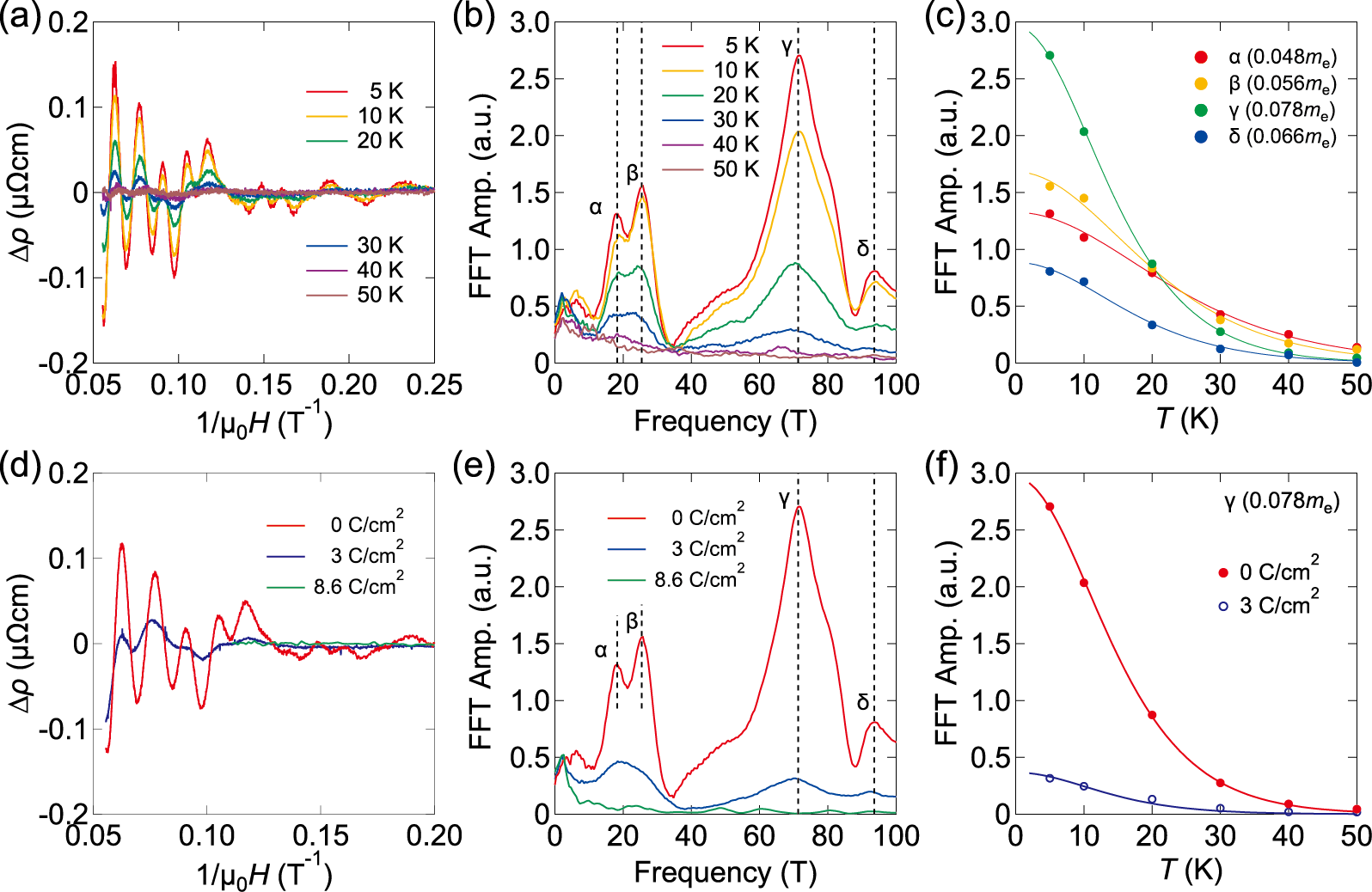}
    \caption{(a) Oscillatory components extracted from the MR of the pristine sample at various temperatures. (b) FFT spectrum of the pristine sample at various temperatures. (c) Temperature dependence of the FFT amplitude for each orbit. The solid lines represent the fits to the Lifshitz-Kosevich formula. Here, $m_{\rm e}$ is the free electron mass. (d) Oscillatory components extracted from the MR of the pristine and irradiated samples measured at 5\,K. (e) FFT spectrum of the pristine and irradiated samples at 5\,K. (f) FFT amplitudes of the $\gamma$ orbit for the pristine and 3\,C/cm$^2$ irradiated samples. For both samples, the effective mass obtained using the LK formula is $m_{\rm eff}=0.078 m_{\rm e}$.}
    \label{Fig2}
\end{figure*}

First, we analyze the SdH quantum oscillations. Generally, the MR is fitted with a quadratic function to extract the oscillatory components. However, as shown in Fig.\,\ref{Fig1}(a), the MR of CsV$_3$Sb$_5$ exhibits a non-quadratic dependence in the low-field region around 1--2\,T, preventing a straightforward fit across the entire field range. Then,  to obtain the non-oscillatory component of $\rho_{xx}^{\rm bg}$, we simultaneously fit $\sigma_{xx}(H)$ and $\sigma_{xy}(H)$ using generalized Lorentzian functions based on the $\mu$-spectrum analysis (see Supplemental Material).
As a result, we obtain the non-oscillatory components of $\sigma_{xx}^{\rm bg}$ and $\sigma_{xy}^{\rm bg}$, yielding $\rho^{\rm bg}_{xx}=\sigma_{xx}^{\rm bg}/[(\sigma_{xx}^{\rm bg})^2+(\sigma_{xy}^{\rm bg})^2]$. Figure \ref{Fig2}(a) shows the oscillatory components $\Delta\rho(H)=\rho_{xx}(H)-\rho^{\rm bg}_{xx}(H)$ at various temperatures. Clear quantum oscillations are observed below 50\,K. 
Figure\,\ref{Fig2}(b) displays the fast Fourier transform (FFT) of the oscillatory components in Fig.\,\ref{Fig2}(a), revealing four distinct frequencies at 18\,T, 25\,T, 73\,T, and 92\,T, labeled as $\alpha, \beta, \gamma$, and $\delta$, respectively. 
Each frequency $F$ corresponds to the extremal cross-sectional area $A_F$ of each Fermi surface, as expressed by the Onsager relation $F=(\hbar/2\pi e)A_F$. 
The $A_F$ values obtained for these frequencies are equivalent to $0.0044A_{\mathrm{BZ}}$, $0.0062A_{\mathrm{BZ}}$, $0.018A_{\mathrm{BZ}}$, and $0.023A_{\mathrm{BZ}}$, where $A_{\mathrm{BZ}}$ is the area of the Brillouin zone (BZ) in the CDW phase, indicating the presence of extremely tiny Fermi pockets. 
Figure\,\ref{Fig2}(c) presents the temperature dependence of the FFT amplitude for each Fermi pocket, from which the effective cyclotron mass can be evaluated using the Lifshitz-Kosevich (LK) formula, $a_i=(\lambda m_{\rm eff}T/\mu_0H_{\mathrm{avg}})/\mathrm{sinh}(\lambda m_{\rm eff}T/\mu_0H_\mathrm{avg})$, where $a_i$ is the FFT amplitude, $\lambda=14.69$\,T/K, $m_{\rm eff}$ is the effective cyclotron mass, $\mu_0$ is the vacuum permeability, and $H_\mathrm{avg}$ is the average value of the magnetic field window used for the FFT. All effective masses obtained here are extremely low, consistent with previous studies\,\cite{ortiz2021,yu2021AHE}. 

\begin{figure*}
    \centering
    \includegraphics[width=1\linewidth]{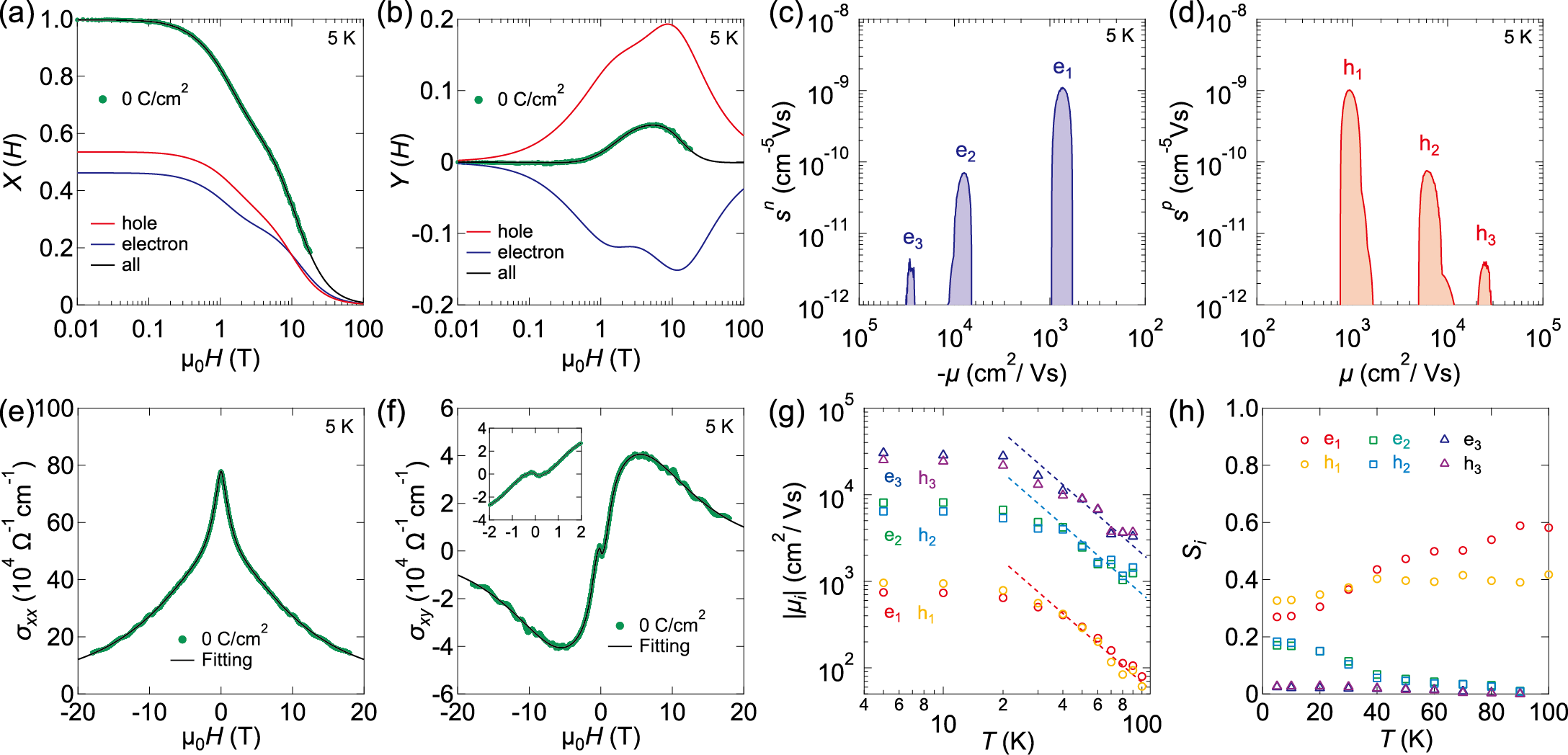}
    \caption{(a), (b) Normalized MC, $X(H)$, and Hall conductivity, $Y(H)$, calculated at 5\,K for pristine CsV$_3$Sb$_5$, respectively. The black lines represent Lorentzian fits to the experimental results (green). Red and blue lines indicate contributions from hole and electron carriers, respectively. (c), (d) Normalized conductivity spectra as a function of mobility for hole and electron carriers in pristine CsV$_3$Sb$_5$ at 5\,K, respectively. (e), (f) Comparisons between experimental and fitting results for $\sigma_{xx}(H)$ and $\sigma_{xy}(H)$, respectively. The inset in (f) shows the expanded view in the low-field region. (g), (h) Temperature dependence of the averaged mobility $|\mu_i|$ of each carrier and the contribution of each carrier to the total conductivity $S_i$. The dashed lines in (g) indicate a $T^{-2}$ dependence.}
    \label{Fig3}
\end{figure*}

Next, we investigate the effects of these small Fermi pockets on the magnetotransport properties of CsV$_3$Sb$_5$. To this end, we employ the $\mu$-spectrum method\,\cite{mcclure1958,Beck1987,Antoszewski2012,Huynh2014}, which allows for the analysis of magnetotransport properties without assumptions regarding the number or types of carriers (see Supplemental Material). For this analysis, we first define the normalized MC, $X(H)\equiv \sigma_{xx}(H)/\sigma_{xx}(0)$, and Hall conductivity, $Y(H)\equiv \sigma_{xy}(H)/\sigma_{xx}(0)$, as shown in Figs.\,\ref{Fig3}(a) and \ref{Fig3}(b). 
Generally, the MC and Hall conductivity can be separated into hole and electron contributions, i.e., $X(H)=X^p(H)+X^n(H)$ and $Y(H)=Y^p(H)+Y^n(H)$, where the superscripts $p$ and $n$ represent the contributions from hole and electron carriers, respectively. 
Additionally, the Kramers-Kronig (KK) relations lead to $X'(H)=Y^p(H)-Y^n(H)$ and $Y'(H)=-X^p(H)+X^n(H)$, where $X'(H)$ and $Y'(H)$ denote the KK transformations of $X(H)$ and $Y(H)$, respectively\,\cite{mcclure1958}\,(see Supplemental Material). Using these relations, we can obtain $X^{p,\,n}(H)$ and $Y^{p,\,n}(H)$, as shown in Figs.\,\ref{Fig3}(a) and \ref{Fig3}(b).

Here, each component of  $X^{p,\,n}$ and $Y^{p,\,n}$ can be represented as an integral of Lorentzian functions with respect to the carrier mobility $\mu$ as follows\,\cite{mcclure1956}: 
\begin{equation}
    X^{p,\,n}(H)=\int_0^\infty \frac{s^{p,\,n}(\mu)d\mu}{1+\mu^2B^2},
\end{equation}
\begin{equation}
    Y^{p,\,n}(H)=\int_0^\infty \frac{s^{p,\,n}(\mu)\mu B d\mu}{1+\mu^2B^2},
\end{equation}
where $s^{p,\,n}(\mu)$ denotes the normalized conductivity as a function of $\mu$, and $B=\mu_0H$ represents the magnetic flux density. 
In this study, to obtain $s^{p,\,n}(\mu)$ as a function of $\mu$, we divide $\mu$ into $N$ segments so that $\ln{(\mu)}$ is equally spaced. We then fit $X_{p,\,n}(H)$ and $Y_{p,\,n}(H)$ for $N$ different magnetic fields, using $s^{p,\,n}(\mu_j)$ as a fitting parameter for each $\mu_j\,(j=1\sim N)$ (see Supplemental Material). Figures\,\ref{Fig3}(c) and \ref{Fig3}(d) show the normalized conductivity spectrum obtained at 5\,K, where positive (negative) mobility corresponds to hole (electron) carriers. The $s^{p,\,n}$ spectrum reveals three distinct peaks for both hole and electron carriers, indicating the presence of three hole carriers ($h_1$, $h_2$, and $h_3$) and three electron carriers ($e_1$, $e_2$, and $e_3$). Subsequently, we calculated the averaged mobility ($\mu_i$) and the contribution of each peak $i\,(=e_1$--$e_3, h_1$--$h_3$) to the total conductivity ($S_i$). Using these values as initial parameters, we fitted $\sigma_{xx}(H)$ and $\sigma_{xy}(H)$ with Lorentzian functions (see Supplemental Material). As a result, as shown in Figs.\,\ref{Fig3}(e) and \ref{Fig3}(f), the fitted MC and Hall conductivity exhibit good agreement with the experimental results.

Figure\,\ref{Fig3}(g) shows the temperature dependence of the mobility $|\mu_i|$ of each carrier in the CDW phase of pristine CsV$_3$Sb$_5$. At high temperatures, the mobility of all carriers follows typical Fermi liquid behavior ($\propto 1/T^2$) but saturates below 30\,K due to impurity scatterings. Notably, the four carriers, $h_2$, $h_3$, $e_2$, and $e_3$, exhibit extremely high mobility ($\gtrsim\sim$10,000\,cm$^2$/Vs) at low temperatures. Given their extremely low effective masses, these carriers may originate from Dirac dispersions.
The comparison between the $\mu$-spectrum analysis and the quantum oscillation results suggests that the $h_2$, $h_3$, $e_2$, and $e_3$ carriers correspond to the $\alpha$--$\delta$ orbits, whereas the $h_1$ and $e_1$ carriers are assigned to larger Fermi surfaces around the $\Gamma$ and K points in the original BZ (see Supplemental Material).
Figure\,\ref{Fig3}(h) shows the temperature dependence of the contribution of each carrier to the total conductivity $S_i$. Although the mobilities of $h_1$ and $e_1$ are not particularly high ($\sim$1,000\,cm$^2$/Vs), their carrier densities are much larger than those of the other four carriers, suggesting that the sign of the Hall effect is determined by the balance between $h_1$ and $e_1$. Indeed, the contributions from $h_1$ and $e_1$ to the total conductivity are reversed at 30\,K, where the sign of the Hall effect is also reversed (see Fig.\,\ref{Fig1}(b)). 
On the other hand, the remaining four carriers, $h_2$, $h_3$, $e_2$, and $e_3$, are characterized by extremely high mobility and low-carrier density, which are expected to play a dominant role in the magnetotransport properties at low temperatures and low magnetic fields.

To examine whether the non-monotonic Hall effect originates from the small Fermi pockets with high-mobility carriers, we investigated the impurity effect on magnetotransport properties of CsV$_3$Sb$_5$, using the electron irradiation technique. It is well known that electron irradiation can create non-magnetic point defects without changing the lattice constants and carrier concentration of the sample\,\cite{Mizukami2014,Roppongi2023}.
To verify that the electronic structure remains unaffected by electron irradiation, we analyzed the SdH quantum oscillations.
Figure\,\ref{Fig2}(d) illustrates the SdH oscillatory components for the samples with irradiated doses of 0 (pristine), 3, and 8.6\,C/cm$^2$. 
The oscillation period is unaffected by electron irradiation, as confirmed by the FFT results of the oscillatory components (see Fig.\,\ref{Fig2}(e)). 
The effective mass of the $\gamma$ orbit for the pristine and 3\,C/cm$^2$ irradiated samples remains unchanged by electron irradiation (see Fig.\,\ref{Fig2}(f)). These results indicate that the Fermi surface structure is essentially unchanged by electron irradiation, and that electron irradiation primarily causes a reduction in the mean free path of the carriers.

Figure\,\ref{Fig1}(e) presents the field dependence of the Hall conductivity for irradiation doses of 0, 3, and 8.6\,C/cm$^2$ measured at 5\,K. With increasing irradiation dose, the non-monotonic behavior of the Hall effect in the low-field region is largely suppressed, accompanied by a shift of the dominant carrier contribution from hole to electron.
The Hall anomaly observed at low fields in this kagome system is often attributed to anomalous Hall mechanisms. Generally, the anomalous Hall effect arising from Berry curvature is dominated by the topological band structure and should be robust against impurities. 
However, we find that although electron irradiation only changes the mobility without affecting the band structure, the non-monotonic Hall component below $\sim$1\,T is significantly suppressed by electron irradiation (see Fig.\,\ref{Fig1}(f)). This fact strongly suggests that the non-monotonic Hall effect at low fields originates not from Berry curvature, but rather from the small Fermi pockets with high mobility. 

In the case of the anomalous Hall effect arising from loop current order, $\sigma_{xy}$ becomes independent of $\sigma_{xx}$ in the moderated scattering regime\,\cite{Tazai2023}. 
However, our results show that $\sigma_{xy}\propto\sigma_{xx}^2$ (see Supplemental Material), which is the typical relationship characteristic of the ordinary Hall effect. Thus, the Hall effect in the present kagome system appears to be governed by the ordinary Hall effect. It is worth noting, however, that our experiments do not rule out the presence of Berry curvature due to loop current order. Generally, the anomalous Hall effect arising from Berry curvature tends to be limited to values around $\sigma_{xy}\sim1,000\,\Omega^{-1}$cm$^{-1}$ (see Supplemental Material).
In contrast, in this kagome system, $\sigma_{xy}$ exhibits a significantly larger value around $10,000\,\Omega^{-1}$cm$^{-1}$ due to the small Fermi pockets with extremely high mobility. Thus, even if the anomalous Hall effect arising from loop current order is present, it should be masked.
Further investigations are required to fully understand the loop current order in this system.

In summary, we performed comprehensive magnetotransport measurements on CsV$_3$Sb$_5$, uncovering the presence of tiny Fermi pockets with extremely high-mobility carriers below the CDW transition. Furthermore, electron irradiation experiments reveal that the non-monotonic Hall effect is significantly suppressed in samples with reduced mobility, despite the absence of substantial changes to the electronic structure. These findings indicate that the non-monotonic Hall effect arises from the tiny Fermi pockets with high mobility rather than the anomalous Hall mechanisms.

\section*{Acknowledgments}
We thank H. Kontani for fruitful discussions. This work was supported by Grants-in-Aid for Scientific Research (KAKENHI) (Nos.\ JP24K17007, JP24H01646, JP23H00089, JP22H00105, JP22H01176, JP22H04933, JP22H00109), Grant-in-Aid for Scientific Research on innovative areas ``Quantum Liquid Crystals" (No.\ JP19H05824), Grant-in-Aid for Scientific Research for Transformative Research Areas (A) ``Condensed Conjugation" (Nos.\ JP20H05869, JP21H05470) from Japan Society for the Promotion of Science (JSPS), and JST, PRESTO (No.\ JPMJPR2458), Japan.
This work was performed under the GIMRT Program of the Institute for Materials Research, Tohoku University (Proposal Nos.\ 202012-HMKPA-0057, 202112-HMKPA-0018, 202212-HMKPA-0055).
Electron irradiation experiments performed at the SIRIUS beamline were supported by the EMIR\&A French network (FR CNRS 3618).
Work by B.R.O. is supported by the U.S. Department of Energy, Office of Science, Basic Energy Sciences, Materials Sciences and Engineering Division. B.R.O and S.D.W gratefully acknowledge support via the UC Santa Barbara NSF Quantum Foundry funded via the Q-AMASE-i program under award No.\ DMR-1906325.

\bibliography{ref}

\end{document}


\begin{center}
	\textbf{\Large Supplemental Material for}\\
	\vspace{10pt}
		\textbf{\large Impact of tiny Fermi pockets with extremely high mobility on the Hall anomaly in the kagome metal CsV$_3$Sb$_5$}\\
		\vspace{10pt}
S.~Liu$^{\ast}$, M.~Roppongi, M.~Kimata, K.~Ishihara, R.~Grasset, M.~Konczykowski, B.~R.~Ortiz, S.~D.~Wilson, K.~Yoshimi, T.~Shibauchi$^{\dag}$, and K.~Hashimoto$^{\ddag}$\\

$^{\ast}$\,s.liu@qpm.k.u-tokyo.ac.jp\\
$^{\dag}$\,shibauchi@k.u-tokyo.ac.jp\\
$^{\ddag}$\,k.hashimoto@edu.k.u-tokyo.ac.jp
\end{center}

\noindent
\textbf{This PDF file includes:}\\
Supplementary Text\\
Figures\,S1 to S6\\
Table\,S1\\
Supplementary References

\clearpage

\section{Mangetotransport measurements}
\begin{figure}[b]
	\centering
	\includegraphics[width=0.98\linewidth]{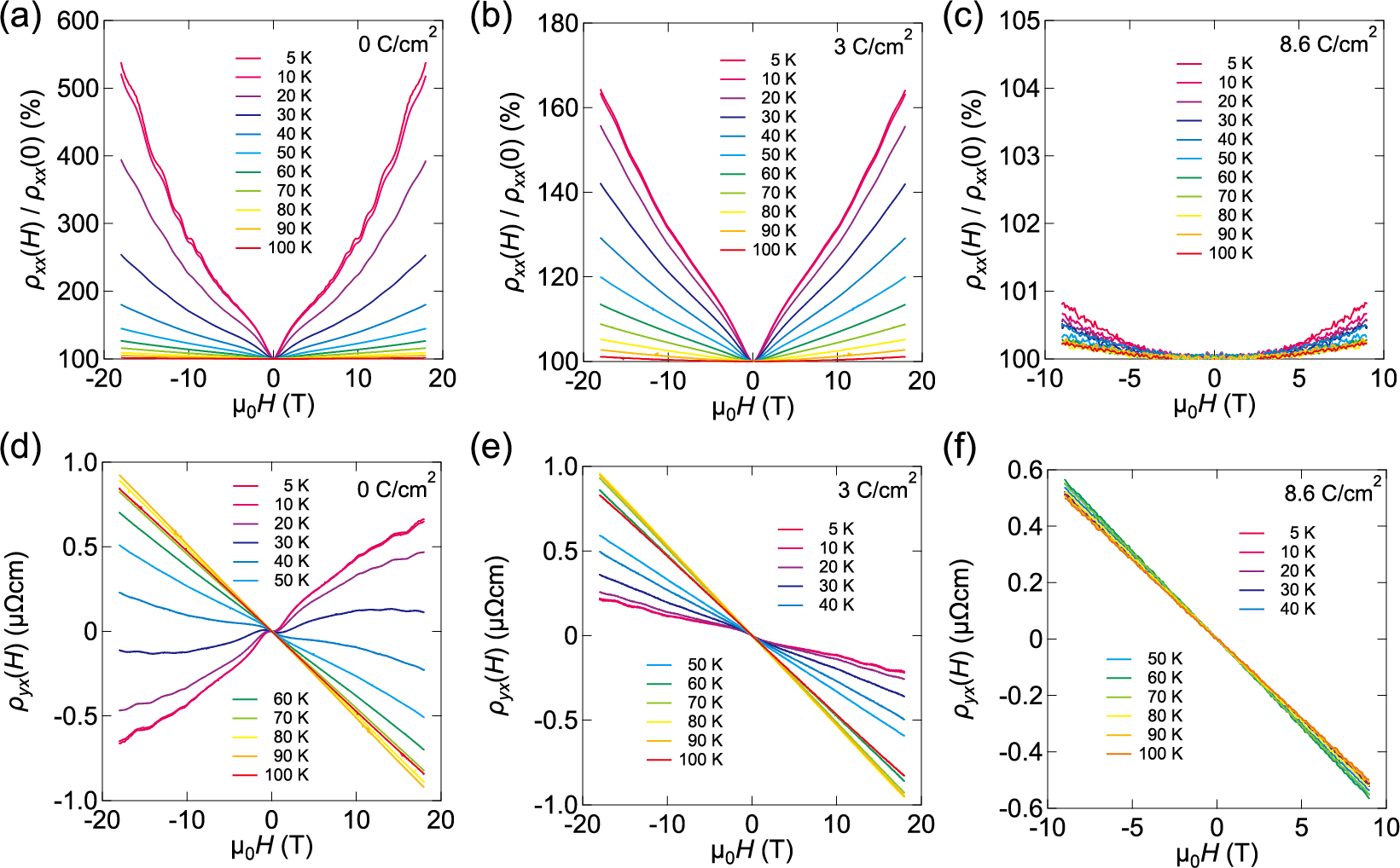}
        \caption{(a)-(c) Normalized mangetoresistance $\rho_{xx}(H)/\rho_{xx}(0)$ of (a) pristine (0\,C/cm$^2$), (b) 3\,C/cm$^2$, and (c) 8.6\,C/cm$^2$ irradiated CsV$_3$Sb$_5$ samples as a function of magnetic field measured at several temperatures. (d)-(f) Hall resistance $\rho_{yx}(H)$ of (d) pristine (0\,C/cm$^2$), (e) 3\,C/cm$^2$, and (f) 8.6\,C/cm$^2$ irradiated CsV$_3$Sb$_5$ samples as a function of magnetic field measured at several temperatures.}
	\label{FS1}
\end{figure}
We measured the raw data $\rho^{\rm raw}_{xx}(H)$ and $\rho^{\rm raw}_{yx}(H)$ in both positive and negative magnetic fields and symmetrized the data using the following formula:
\begin{equation}
\rho_{xx}(H)=\frac{1}{2}(\rho^{\rm raw}_{xx}(H)+\rho^{\rm raw}_{xx}(-H)),
\end{equation}
\begin{equation}
\rho_{yx}(H)=\frac{1}{2}(\rho^{\rm raw}_{yx}(H)-\rho^{\rm raw}_{yx}(-H)).
\end{equation}
Figure\,S1 presents $\rho_{xx}(H)/\rho_{xx}(0)$ and $\rho_{yx}(H)$ as a function of magnetic field for pristine (0\,C/cm$^2$), 3\,C/cm$^2$, and 8.6\,C/cm$^2$ electron-irradiated CsV$_3$Sb$_5$ samples.
As the irradiation dose increases, the increase in magnetoresistance against magnetic fields is significantly suppressed, and the pronounced temperature dependence of the Hall resistivity becomes negligibly small.

\clearpage

\section{$\mu$-spectrum analysis}
In this study, we employed the $\mu$-spectrum method\,\cite{mcclure1958,Beck1987,Antoszewski2012,Huynh2014} to extract the conductivity spectrum as a function of mobility $\mu$ from magnetotransport measurements. The detailed analysis process is illustrated in Fig.\,S2 and is described in the following subsections.

\begin{figure}[h]
	\centering
	\includegraphics[width=1\linewidth]{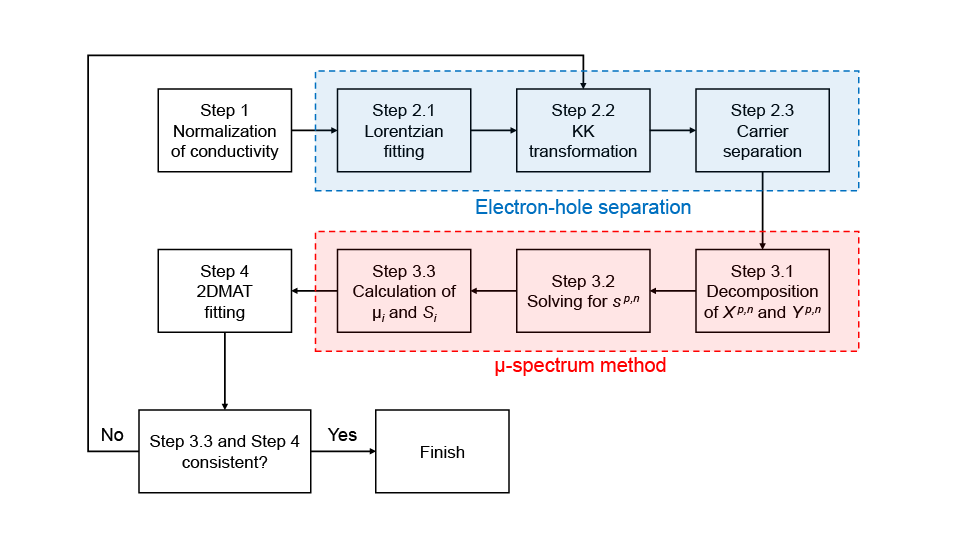}
        \caption{Overview of the analysis process using the $\mu$-spectrum method.}
	\label{FS2}
\end{figure}

\noindent
\subsection*{1. Normalization of conductivity}
The first step is to derive the magnetoconductivity $\sigma_{xx}(H)$ and Hall conductivity $\sigma_{xy}(H)$ from the measured magnetoresistivity $\rho_{xx}(H)$ and Hall resistivity $\rho_{yx}(H)$ using the following relations:
\begin{equation}
\sigma_{xx}(H) = \frac{\rho_{xx}(H)}{\rho_{xx}^2(H) + \rho_{yx}^2(H)},
\end{equation}
\begin{equation}
\sigma_{xy}(H) = \frac{\rho_{yx}(H)}{\rho_{xx}^2(H) + \rho_{yx}^2(H)}.
\end{equation}
To facilitate further analysis, we normalize $\sigma_{xx}(H)$ and $\sigma_{xy}(H)$ by the zero-field conductivity $\sigma_{xx}(0)$, defining the dimensionless magnetoconductivity $X(H)$ and Hall conductivity $Y(H)$ as follows:
\begin{equation}
X(H) = \frac{\sigma_{xx}(H)}{\sigma_{xx}(0)},
\end{equation}
\begin{equation}
Y(H) = \frac{\sigma_{xy}(H)}{\sigma_{xx}(0)}.
\end{equation}
As discussed in the next  subsection, these normalized quantities are essential for performing the Kramers-Kronig (KK) transformation to separate the electron and hole carrier contributions to the conductivity.

\noindent
\subsection*{2. Electron-hole separation}
To obtain the mobility spectrum from the magnetotransport data, it is essential to separate the contributions of electron and hole carriers. Due to the multi-band nature of the present system, the electron and hole contributions in $X(H)$ and $Y(H)$ partially cancel each other, leading to ambiguity in the analysis process. To overcome this issue, we employ a three-step procedure: (i) Lorentzian fitting, (ii) KK transformation, and (iii) carrier separation. Each step is described in detail below.
 
\subsection*{2.1 Lorentzian fitting}
In general, $X(H)$ and $Y(H)$ can be expressed as a sum of Lorentzian functions with respect to the magnetic field. 
As will be discussed below, the KK transformation of $X(H)$ and $Y(H)$ requires knowledge of their values across the entire range of $H$. To achieve this, we fit the experimental data using a linear combination of Lorentzian functions, allowing us to determine their values in the high-field region. The fitting functions are given by
\begin{equation}
X(H) = \sum_{j = 1}^{n} \frac{a_j}{1 + \mu_j^2 B^2},
\end{equation}
\begin{equation}
Y(H) = \sum_{j = 1}^{n} \frac{b_j B}{1 + \mu_j^2 B^2},
\end{equation}
where $n$ is the total number of Lorentzian functions, $a_j$, $b_j$, and $\mu_j$ are fitting parameters for the $j$-th Lorentzian function, and $B = \mu_0 H$ is the magnetic flux density.

In our initial analysis, we found that choosing $n = 4$ or $5$ provided a reasonable fit to the experimental data. However, the exact value of $n$ does not carry significant physical meaning. The primary purpose of this fitting procedure is to ensure that $X(H)$ and $Y(H)$ have well-defined values over the entire magnetic field range, which is necessary for performing the KK transformation. The fitting results using the sum of the Lorentzian functions described above are shown in Figs.\,S3(a) and S3(b).

\begin{figure}[h] 
    \centering
    \includegraphics[width=0.85\textwidth]{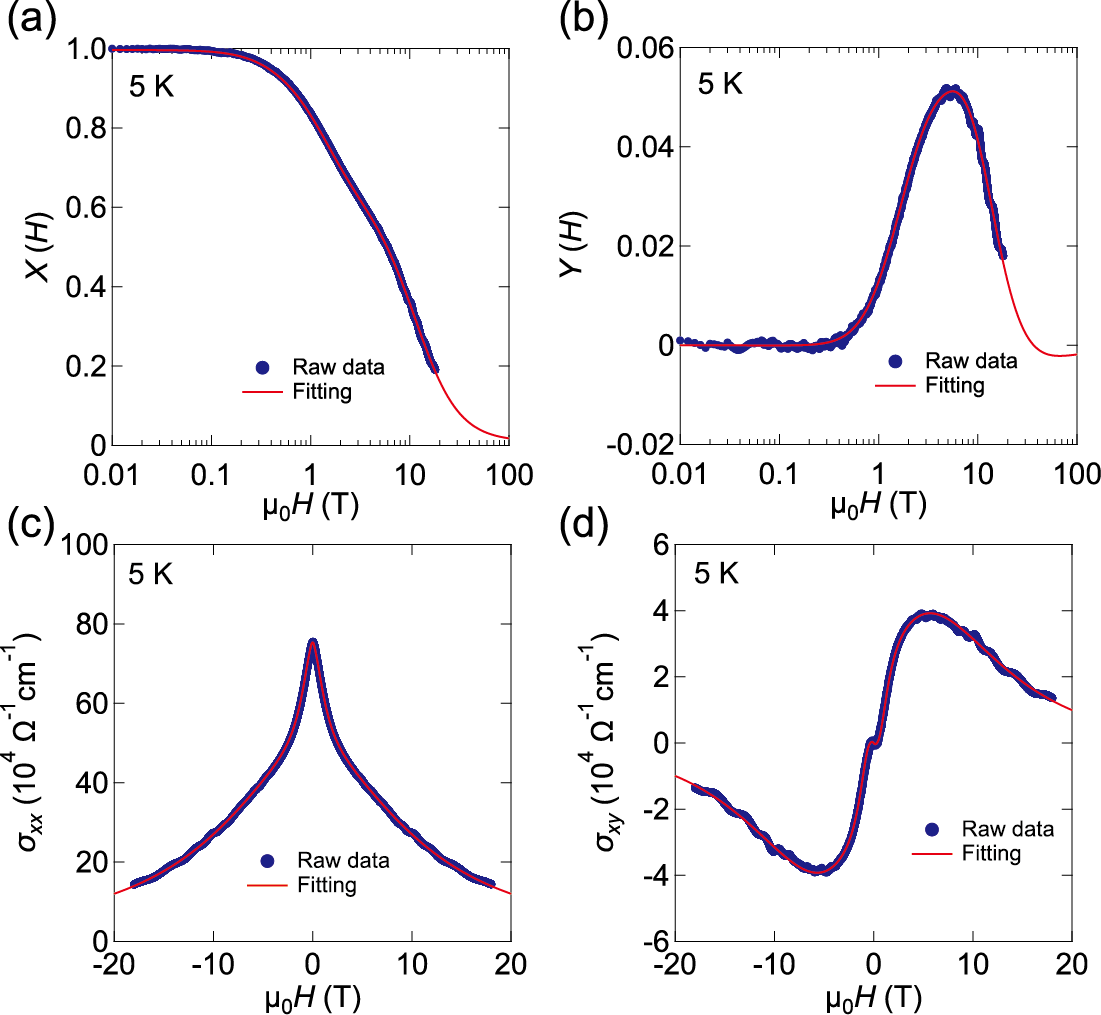}
    \caption{(a), (b) $X(H)$ and $Y(H)$ at 5\,K for pristine CsV$_3$Sb$_5$, respectively. The red lines represent Lorentzian fits to the experimental results (blue). (c), (d) Comparison between experimental and fitting results for $\sigma_{xx}(H)$ and $\sigma_{xy}(H)$, respectively.}
    \label{FS3}
\end{figure}

\subsection*{2.2 Kramers-Kronig (KK) transformation}
To separate the contributions of electron and hole carriers, we apply the KK transformation to $X(H)$ and $Y(H)$ using the following expressions:
\begin{equation}
X'(H) = \mathcal{P} \frac{1}{\pi} \int_{-\infty}^{\infty} \frac{X(H')}{\mu_0 H - \mu_0 H'} dH' = Y^p(H) - Y^n(H),
\label{KK1}
\end{equation}
\begin{equation}
Y'(H) = \mathcal{P} \frac{1}{\pi} \int_{-\infty}^{\infty} \frac{Y(H')}{\mu_0 H - \mu_0 H'} dH' = -X^p(H) + X^n(H),
\label{KK2}
\end{equation}
where $\mathcal{P}$ denotes the Cauchy principal value, and the superscripts $p$ and $n$ indicate contributions from hole and electron carriers, respectively. Equations\,(\ref{KK1}) and (\ref{KK2}) show that $X'(H)$ and $Y'(H)$ can be divided into the hole and electron carrier contributions.
In practical calculations, the integration range is inevitably finite. Therefore, in this study, to perform the KK transformation accurately, we set the integration range from $-10^6$\,T to $10^6$\,T so that $X(H)$ and $Y(H)$ approach zero within the integration range. This minimizes numerical artifacts and improves the accuracy of the KK transformation.

\subsection*{2.3 Carrier separation}  
In general, the following relations hold for $X(H)$ and $Y(H)$:
\begin{equation}
X(H) = X^p(H) + X^n(H),
\end{equation}
\begin{equation}
Y(H) = Y^p(H) + Y^n(H).
\end{equation}
Using these equations along with Eqs.\,(\ref{KK1}) and (\ref{KK2}), we obtain explicit expressions for $X^p(H)$, $X^n(H)$, $Y^p(H)$, and $Y^n(H)$ as follows:
\begin{equation}
X^p(H) = \frac{1}{2} \left[ X(H) - Y'(H) \right],
\end{equation}
\begin{equation}
X^n(H) = \frac{1}{2} \left[ X(H) + Y'(H) \right],
\end{equation}
\begin{equation}
Y^p(H) = \frac{1}{2} \left[ Y(H) + X'(H) \right],
\end{equation}
\begin{equation}
Y^n(H) = \frac{1}{2} \left[ Y(H) - X'(H) \right].
\end{equation}
These relations allow us to separately evaluate the contributions of hole and electron carriers to the conductivity. The obtained results are presented in Figs.\,3(a) and (b) in the main text.

\subsection*{3. $\mu$-spectrum method}
\subsection*{3.1 Decomposition of $X^{p,\,n}(H)$ and $Y^{p,\,n}(H)$}
In the first step of the $\mu$-spectrum method, we decompose $X^{p,\,n}(H)$ and $Y^{p,\,n}(H)$ into a sum of Lorentzian functions. In general, as discussed in the main text, $X^{p,\,n}(H)$ and $Y^{p,\,n}(H)$ can be expressed as an integral of Lorentzian functions with respect to the carrier mobility $\mu$ (see Eqs.\,(1) and (2) in the main text). If $s^{p,\,n}(\mu)$ can be determined as a function of $\mu$, the conductivity-mobility spectrum can be obtained.
To achieve this, we discretize the integral expression by introducing a small interval $\Delta \mu$. However, choosing $\Delta \mu$ with an equal spacing would lead to a large ambiguity in the low-field region. Instead, we define the discretization such that $\ln(\mu)$ is evenly spaced, i.e., we set the interval as $\Delta m = \Delta \ln(\mu)$. Using this formulation, $X^{p,\,n}(H)$ and $Y^{p,\,n}(H)$ are expressed as a sum of Lorentzian functions as follows:
\begin{equation}
X^{p,\,n}(H) = \sum_{j = 1}^{N} \frac{\exp{(m_j)} s^{p,\,n}(m_j) \Delta m}{1 + \exp[2(m_j + b)]},
\end{equation}
\begin{equation}
Y^{p,\,n}(H) = \sum_{j = 1}^{N} \frac{\exp{(m_j)} s^{p,\,n}(m_j) \Delta m}{2 \mathrm{cosh}(m_j + b)},
\end{equation}
where $m = \mathrm{ln}(\mu)$, $b = \mathrm{ln}(B)$, $N$ represents the number of divisions used for discretization, and $s^{p,\,n}(\mu)$ is the conductivity-mobility spectrum. 
Here, $m$ is divided into $N=1000$ segments, ranging from $-6$ to 4 with an interval of $\Delta m=0.01$. Similarly, $b$ is also discretized into  1000 points, ranging from $-6$ to $4$ with an interval of $\Delta b=0.01$. 
This discretization allows $X^{p,\,n}(H)$ and $Y^{p,\,n}(H)$ to be expressed in matrix form as follows:
\begin{equation}
\begin{pmatrix}
    X^{p,\,n}(b_{1}) \\
    X^{p,\,n}(b_{2}) \\
    \vdots \\
    X^{p,\,n}(b_{N})
\end{pmatrix} 
=
\begin{pmatrix}
    \frac{\exp{(m_1)} \Delta m}{1+\mathrm{exp}[2(m_1+b_1)]} & \cdots & \frac{\exp{(m_N)} \Delta m}{1+\mathrm{exp}[2(m_{N}+b_1)]} \\
    \frac{\exp{(m_1)} \Delta m}{1+\mathrm{exp}[2(m_1+b_2)]} & \cdots & \frac{\exp{(m_N)} \Delta m}{1+\mathrm{exp}[2(m_{N}+b_2)]} \\
    \vdots & \ddots & \vdots \\
    \frac{\exp{(m_1)} \Delta m}{1+\mathrm{exp}[2(m_1+b_{N})]} & \cdots & \frac{\exp{(m_N)} \Delta m}{1+\mathrm{exp}[2(m_{N}+b_{N})]}
\end{pmatrix}	
\begin{pmatrix}
    s^{p,\,n}(m_{1}) \\
    s^{p,\,n}(m_{2}) \\
    \vdots \\
    s^{p,\,n}(m_{N})
\end{pmatrix},
\label{matrix1}
\end{equation}
\begin{equation}
\begin{pmatrix}
    Y^{p,\,n}(b_{1}) \\
    Y^{p,\,n}(b_{2}) \\
    \vdots \\
    Y^{p,\,n}(b_{N})
\end{pmatrix} 
=
\begin{pmatrix}
    \frac{\exp{(m_1)} \Delta m}{2\mathrm{cosh}(m_1+b_1)} & \cdots & \frac{\exp{(m_N)} \Delta m}{2\mathrm{cosh}(m_{N}+b_1)} \\
    \frac{\exp{(m_1)} \Delta m}{2\mathrm{cosh}(m_1+b_2)} & \cdots & \frac{\exp{(m_N)} \Delta m}{2\mathrm{cosh}(m_{N}+b_2)} \\
    \vdots & \ddots & \vdots \\
    \frac{\exp{(m_1)} \Delta m}{2\mathrm{cosh}(m_1+b_{N})} & \cdots & \frac{\exp{(m_N)} \Delta m}{2\mathrm{cosh}(m_{N}+b_{N})}
\end{pmatrix}	
\begin{pmatrix}
    s^{p,\,n}(m_{1}) \\
    s^{p,\,n}(m_{2}) \\
    \vdots \\
    s^{p,\,n}(m_{N})
\end{pmatrix}.
\label{matrix2}
\end{equation}

\subsection*{3.2 Solving for $s^{p,\,n}$}

In the second step of the $\mu$-spectrum method, we determine $s^{p,\,n}(m_j)$ for each $m_j$ through Eqs.\,(\ref{matrix1}) and (\ref{matrix2}) obtained in the previous step. To achieve this, we substitute numerical values for $b_j$ and $m_j$ within the range of $-6$ to 4 and solve the matrix equations to obtain $s^{p,\,n}(m_j)$.
In general, to obtain a set of $s^{p,\,n}(m_j)$, it is necessary to compute the inverse of an $N \times N$ matrix. However, the inverse matrix for large $N$ generally involves uncertainty. 
Then, in this study, we search for $s^{p,\,n}$ that minimizes the difference between the left-hand and right-hand sides of Eqs.\,(\ref{matrix1}) and (\ref{matrix2}).
To this end, first, we set all initial parameters of $s^{p,\,n}(m_j)$ to zero and optimized $s^{p,\,n}(m_j)$ for $X^{p,\,n}(H)$. The optimize values are then used as the initial parameters for $Y^{p,\,n}(H)$. This process is repeated, alternating between $X^{p,\,n}(H)$ and $Y^{p,\,n}(H)$, until the obtained $s^{p,\,n}(m_j)$ converged to consistent values, because $X^{p,\,n}(H)$ and $Y^{p,\,n}(H)$ should be described by the same set of $s^{p,\,n}(m_j)$.
For computational efficiency, we utilized a GPU-based fitting package\,\cite{gpufit}. The obtained conductivity-mobility spectrum is presented in Figs.\,3(c) and 3(d) in the main text.

\subsection*{3.3 Calculation of $\mu_i$ and $S_i$}
In the final step of the $\mu$-spectrum method, we calculate the averaged mobility for each carrier and the contribution of each carrier to the total conductivity. The obtained mobility spectrum exhibits three characteristic peak structures for both electron and hole carriers (see Figs.\,3(c) and 3(d) in the main text). To determine the mobility and conductivity associated with each carrier, it is crucial to define these quantities properly, given the broad nature of the peaks. To address this, we define the averaged mobility ($\mu_i$) and the contribution of each peak $i\,(=h_1$--$h_3, e_1$--$e_3$) to the total conductivity ($S_i$) using the following expressions:
\begin{equation}
\mu_i = \frac{\sum^{\mu+\varepsilon}_{\mu-\varepsilon} \mu s(\mu) \Delta\mu}{\sum^{\mu+\varepsilon}_{\mu-\varepsilon} s(\mu)\Delta\mu},
\end{equation}
\begin{equation}
S_i = \sum^{\mu+\varepsilon}_{\mu-\varepsilon} s(\mu) \Delta \mu,
\end{equation}
where the summations are taken over each peak $i$ with the peak width of $2\epsilon$. The extracted results are presented in Figs.\,3(g) and 3(h) in the main text. Through the $\mu$-spectrum method, we obtained the following relationship between the averaged mobility $\mu_i$ and the carrier density $n_i$ for each carrier using $S_i$:
\begin{equation}
    \sigma_{xx}(0) S_i = n_i e \mu_i .
\end{equation}

\subsection*{4. 2DMAT fitting}
To obtain more accurate mobility and carrier density, we fit $\sigma_{xx}(H)$ and $\sigma_{xy}(H)$ simultaneously using the values obtained from the $\mu$-spectrum method as initial parameters. This procedure allows us to refine the mobility and carrier density for each carrier (see Figs.\,S3(c) and S3(d)). We applied the 2DMAT program\,\cite{2DMAT} to simultaneously fit both $\sigma_{xx}$ and $\sigma_{xy}$. The expressions used for fitting $\sigma_{xx}$ and $\sigma_{xy}$ are given by:
\begin{equation}
\sigma_{xx}(H) = \sum_{j = 1}^{3} \frac{n_{h_j} e \mu_{h_j}}{1 + \mu_{h_j}^2 B^2} - \sum_{j = 1}^{3} \frac{n_{e_j} e \mu_{e_j}}{1 + \mu_{e_j}^2 B^2},
\end{equation}
\begin{equation}
\sigma_{xy}(H) =  \sum_{j = 1}^{3} \frac{n_{h_j} e \mu_{h_j}^2 B}{1 + \mu_{h_j}^2 B^2} -\sum_{j = 1}^{3} \frac{n_{e_j} e \mu_{e_j}^2 B}{1 + \mu_{e_j}^2 B^2} .
\end{equation}
The optimized parameters obtained from 2DMAT fitting are then used as new initial values to restart the iterative process from the Lorentzian fitting in step\,2 (see Fig.\,S2). The entire sequence of calculations is repeated until the extracted mobility and carrier density converged. Once convergence is achieved, the final mobility and carrier density are determined, completing the analysis.

\clearpage
\section{Quantum oscillation analysis}

Quantum oscillations offer valuable insights into the electronic structure of materials. By analyzing the oscillatory components of transport measurements in a magnetic field, we can extract key parameters such as the Fermi surface topology, effective mass, and carrier scattering rates. In this section, we present the quantum oscillation analysis conducted on the 3\,C/cm$^2$ irradiated sample. Figure\,S4(a) shows the quantum oscillation components of the magnetoresistance for the 3\,C/cm$^2$ irradiated CsV$_3$Sb$_5$ sample. Figure\,S4(b) presents the fast Fourier transform (FFT) of the oscillatory components, revealing peaks at approximately 20\,T, 70\,T, and 90\,T, which are labeled as $\alpha/\beta$, $\gamma$, and $\delta$, respectively. 
Figure\,S4(c) displays the temperature dependence of the FFT amplitude for each orbit, from which the effective mass can be extracted. Notably, electron irradiation almost does not alter the frequency and effective mass of each orbit, indicating that the electronic structure remains unaffected by electron irradiation. Instead, only the carrier scattering rate is changed.
Figure\,S4(d) presents the Landau fan diagram for both the pristine and 3\,C/cm$^2$ irradiated samples. As reported in previous studies, the $\alpha$ and $\gamma$ orbits exhibit a nontrivial Berry phase, whereas the Berry phases of the $\beta$ and $\delta$ orbits are nearly zero. This result suggests that the $\alpha$ and $\gamma$ orbits may be associated with Dirac bands. It should be noted that even if a nontrivial band topology with a finite Berry curvature is present, its contribution to the Hall conductivity is at most about 1,000\,$\Omega^{-1}$cm$^{-1}$, as discussed in other anomalous Hall systems\,\cite{Miyasato2007,iguchi2007,kanazawa2011,Nakatsuji2015,Nayak2016,Ye2018,Liu2018,kurumaji2019,Fujishiro2021} (see Fig.\,S6). 
Therefore, it is unlikely that the non-monotonic Hall effect observed in CsV$_3$Sb$_5$ comes from the non-trivial band topology.
Furthermore, electron irradiation of 3\,C/cm$^2$ does not change the Berry phase of the $\gamma$ orbit, while the non-monotonic Hall effect is no longer observed after electron irradiation.
This also confirms that the nontrivial band topology is not related to the hole anomaly observed in CsV$_3$Sb$_5$.

\begin{figure}[t]
	\centering
	\includegraphics[width=0.75\linewidth]{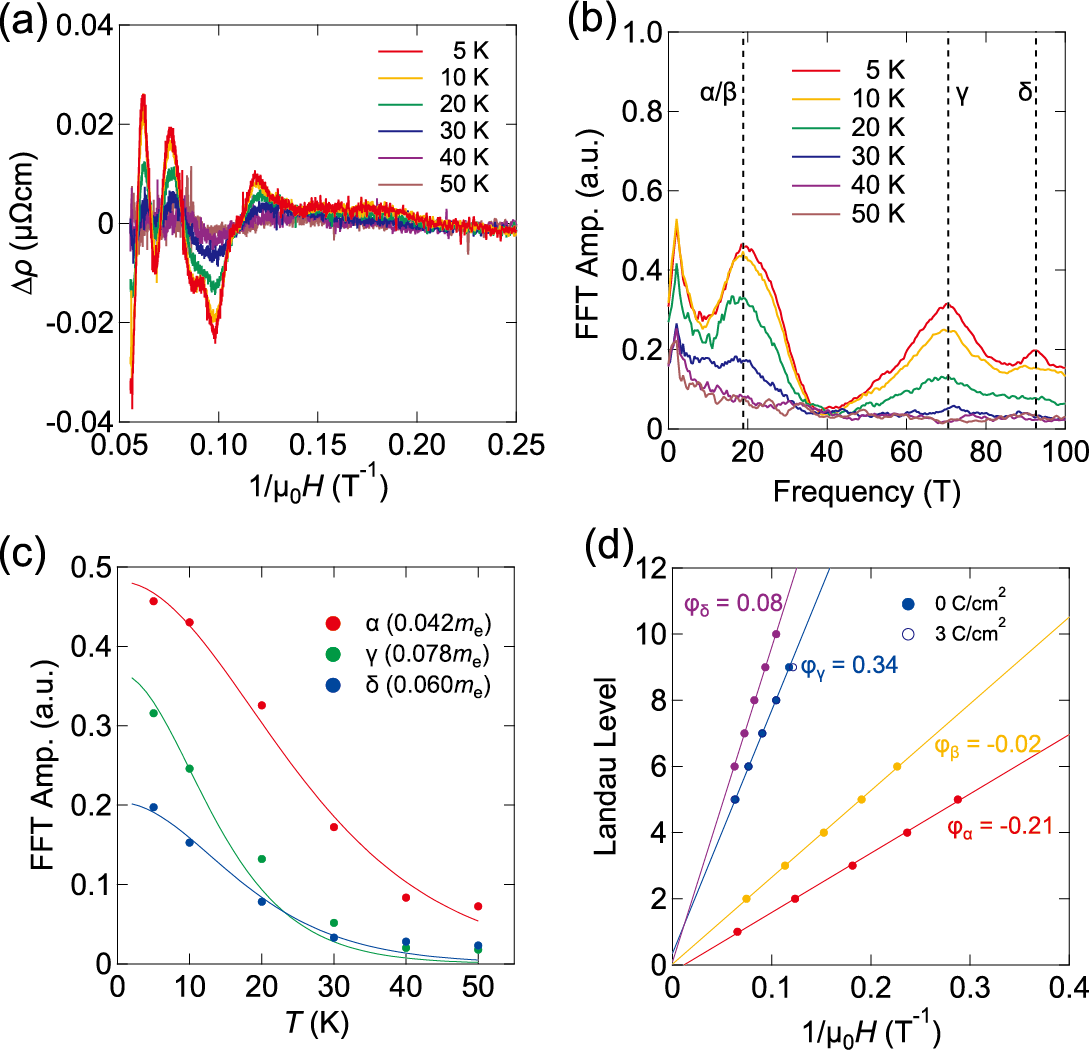}
        \caption{(a) Oscillatory components extracted from the magnetoresistivity for the 3\,C/cm$^2$ irradiated sample at various temperatures. (b) FFT spectrum of the 3\,C/cm$^2$ irradiated sample at various temperatures. Three peaks ($\alpha/\beta$, $\gamma$, and $\delta$ orbits) are observed, in addition to the first peak at the lowest frequency ($\sim$3\,T), which is an artifact of the FFT. (c) Temperature dependence of FFT amplitude for each peak. The solid lines represent the Lifshitz-Kosevich formula. (d) Landau fan diagram plot of the pristine and 3\,C/cm$^2$ irradiated CsV$_3$Sb$_5$ samples.}
	\label{FS4}
\end{figure}

\clearpage
\section{Comparison between quantum oscillation experiments and $\mu$-spectrum analysis}

We compare the results of quantum oscillation measurements and $\mu$-spectrum analysis. Table\,\ref{comparison} summarizes the correspondence between the carriers identified by these two methods. The peaks $h_2, h_3, e_2,$ and $e_3$ identified in the $\mu$-spectrum analysis exhibit low carrier densities, which are likely related to the low-frequency orbits $\alpha, \beta, \gamma$, and $\delta$ detected in the quantum oscillations. These results suggest that they originate from small Fermi pockets below the CDW transition.
The carrier densities estimated from the two methods differ by a factor of several. This discrepancy may arise because quantum oscillation measurements provide the carrier density for a single orbit, whereas in $\mu$-spectrum analysis, the estimated carrier density includes contributions from multiple identical Fermi surfaces within the BZ.
The peaks $h_1$ and $e_1$ are considered to correspond to the high-frequency branches $\mu$ and $\nu$ reported in a previous quantum oscillation study\,\cite{ortiz2021}, which are associated with the Fermi surfaces around the $\Gamma$ and K points in the original BZ. 
Band structure calculations indicate that these Fermi surfaces persist without significant reconstruction across the CDW transition.
The observation of peaks $h_1$ and $e_1$ in the $\mu$-spectrum analysis above the CDW transition supports the validity of our present arguments.

\begin{table}[h]
\centering
\caption{Comparisons between the quantum oscillation experiments and $\mu$-spectrum analysis. The frequency $F$ and carrier density $n$ of each branch observed in the quantum oscillation experiments are listed, where $n$ corresponds to a single Fermi surface orbit. The peaks and corresponding carrier densities obtained from the $\mu$-spectrum analysis are also listed. 
\\
}
\label{comparison}
\begin{tabular}{ccc|cc}
\hline
\multicolumn{3}{c}{Quantum oscillations} & \multicolumn{2}{|c}{$\mu$-spectrum analysis} \\ \hline
\multicolumn{1}{c}{Branch} & \multicolumn{1}{|c|}{$F$\,(T)} & $n$\,(cm$^{-3}$) & \multicolumn{1}{c|}{Peak} & $n$\,(cm$^{-3}$) \\ \hline
\multicolumn{1}{c|}{$\alpha$} & \multicolumn{1}{c|}{18} & $9.35 \times 10^{18}$ & \multicolumn{1}{c|}{$h_3$} & $(2.58 \pm 1.25) \times 10^{18}$ \\ 
\multicolumn{1}{c|}{$\beta$} & \multicolumn{1}{c|}{25} & $1.30 \times 10^{19}$ & \multicolumn{1}{c|}{$e_3$} & $(4.26 \pm 0.31) \times 10^{19}$ \\ 
\multicolumn{1}{c|}{$\gamma$} & \multicolumn{1}{c|}{73} & $3.79 \times 10^{19}$ & \multicolumn{1}{c|}{$h_2$} & $(1.29 \pm 0.47) \times 10^{20}$ \\ 
\multicolumn{1}{c|}{$\delta$} & \multicolumn{1}{c|}{93} & $4.83 \times 10^{19}$ & \multicolumn{1}{c|}{$e_2$} & $(0.84 \pm 0.36) \times 10^{20}$ \\ \hline
\multicolumn{1}{c|}{$\mu$\,\cite{ortiz2021}} & \multicolumn{1}{c|}{1700} & $8.83 \times 10^{20}$ & \multicolumn{1}{c|}{$h_1$} & $(1.64 \pm 0.52) \times 10^{21}$ \\ 
\multicolumn{1}{c|}{$\nu$\,\cite{ortiz2021}} & \multicolumn{1}{c|}{1930} & $1.00 \times 10^{21}$ & \multicolumn{1}{c|}{$e_1$} & $(1.75 \pm 0.48) \times 10^{21}$ \\ \hline
\end{tabular}
\end{table}

\clearpage

\section{Non-monotonic components of Hall effect}
To extract the non-monotonic component of the Hall resistivity $\rho_{yx}^{\rm NM}(H)$, we used the following formula:
\begin{equation}
\rho_{yx}^{\rm NM}(H) = \rho_{yx}(H) - c H,
\end{equation}
where $c$ is the coefficient extracted from a linear fit of $\rho_{yx}(H)$ over the magnetic field range of 0--2\,T.
$\rho_{yx}^{\rm NM}(H)$ was then converted into the non-monotonic Hall conductivity $\sigma_{xy}^{\rm NM}(H)$ using the following equation:
\begin{equation}
\sigma_{xy}^{\rm NM}(H) = \frac{\rho_{yx}^{\rm NM}(H)}{\rho_{xx}^2(H) + \rho_{yx}^2(H)}.
\end{equation}
The obtained results for the pristine, 3\,C/cm$^2$, and 8.6\,C/cm$^2$-irradiated samples are presented in Fig.\,S5. 

Here, we compare the obtained non-monotonic Hall conductivity with other materials exhibiting anomalous Hall effects\,\cite{Miyasato2007,iguchi2007,kanazawa2011,Nakatsuji2015,Nayak2016,Ye2018,Liu2018,kurumaji2019,Fujishiro2021} (see Fig.\,S6). Notably, the non-monotonic component of the Hall conductivity in CsV$_3$Sb$_5$ is an order of magnitude larger than those of other materials that exhibit intrinsic anomalous Hall effects. Moreover, we have observed that $\sigma_{xy}^{\rm NM}$ is proportional to $\sigma_{xx}^2$, which is distinctly different from behaviors observed in other anomalous Hall systems.
In CsV$_3$Sb$_5$, the low-field Hall resistivity is much smaller than the magnetoresistivity, allowing the magnetoconductivity and Hall conductivity to be approximated as:
\begin{equation}
\sigma_{xx}(H) = \frac{\rho_{xx}(H)}{\rho_{xx}^2(H) + \rho_{yx}^2(H)} \approx \frac{1}{\rho_{xx}(H)},
\end{equation}
\begin{equation}
\sigma_{xy}(H) = \frac{\rho_{yx}(H)}{\rho_{xx}^2(H) + \rho_{yx}^2(H)} \approx \frac{\rho_{yx}(H)}{\rho_{xx}^2(H)} \approx \rho_{yx}(H) \sigma_{xx}^2(H).
\end{equation}
These relations lead to $\sigma_{xy}\propto\sigma_{xx}^2$, indicating that the non-monotonic component of the Hall conductivity is attributed to the conventional Drude mechanism such as skew scattering.
Furthermore, our $\mu$-spectrum analysis has revealed the presence of extremely high-mobility carriers associated with small Fermi pockets. Considering these findings, we conclude that the non-monotonic Hall effect in CsV$_3$Sb$_5$ does not arise from an anomalous Hall effect but rather originates from high-mobility carriers in small Fermi pockets. Our present findings highlight the importance of high-mobility carriers in the Hall response of CsV$_3$Sb$_5$.

\begin{figure}[t]
	\centering
	\includegraphics[width=1\linewidth]{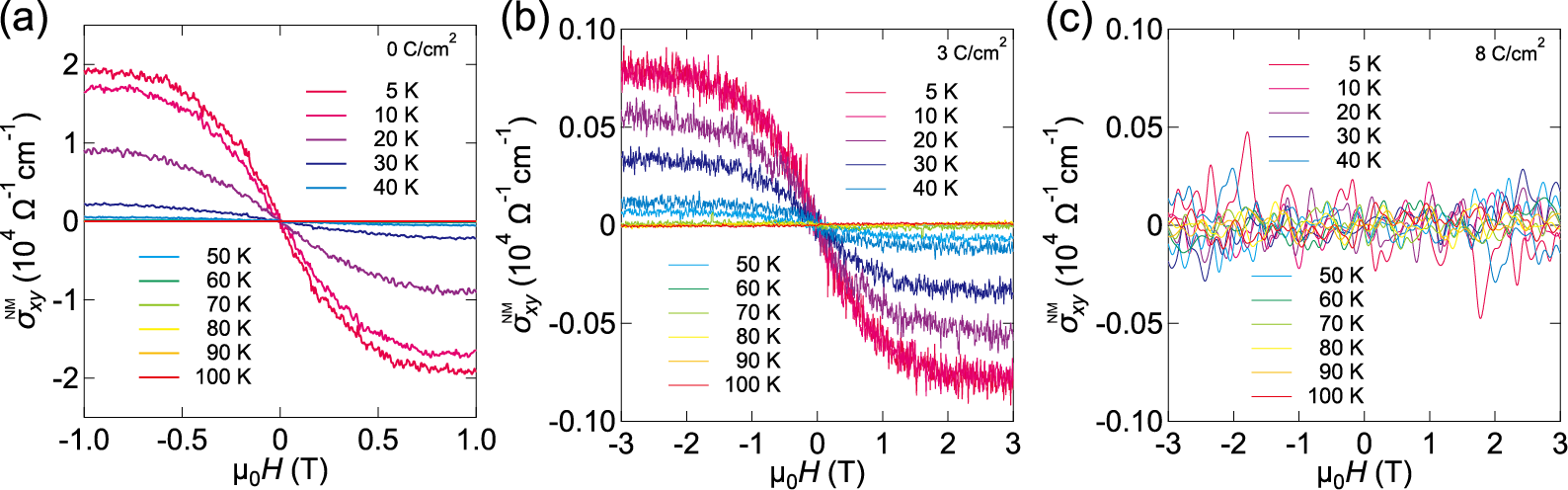}
        \caption{(a)-(c) Non-monotonic components of the Hall conductivity for (a) pristine, (b) 3\,C/cm$^2$, and (c) 8.6\,C/cm$^2$ irradiated CsV$_3$Sb$_5$ samples.}
	\label{FS5}
\end{figure}

\begin{figure}[h]
	\centering
	\includegraphics[width=0.85\linewidth]{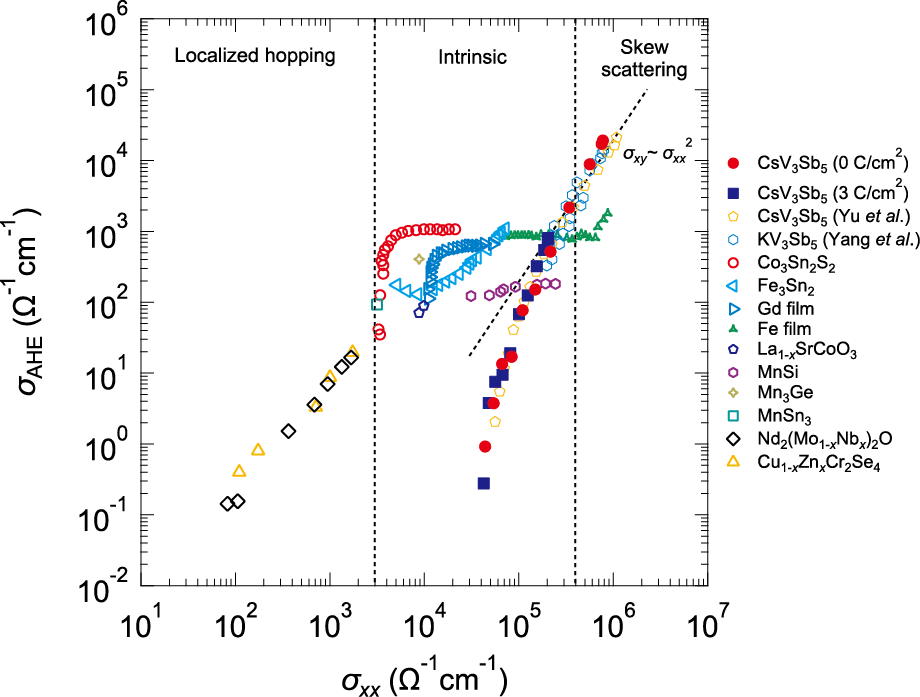}
        \caption{$\sigma_{\rm AHE}$ vs. $\sigma_{xx}$ for a variety of materials. $\sigma^{\rm NM}_{xy}$ vs. $\sigma_{xx}$ for CsV$_3$Sb$_5$ and KV$_3$Sb$_5$, including our present results as well as those from previous studies\,\cite{yu2021AHE,yang2020}, is also plotted. The quadratic scaling between $\sigma_{\rm AHE}$ and $\sigma_{xx}$ is shown as the dashed line.}
	\label{FS6}
\end{figure}

\clearpage
\noindent{\bf Supplementary References}\\
\bibliography{ref_SM.bib}